\documentclass[%
 reprint,
superscriptaddress,
 amsmath,amssymb,
 aps,
]{revtex4-1}

\usepackage{graphicx}% Include figure files
\usepackage{dcolumn}% Align table columns on decimal point
\usepackage{bm}% bold math
\usepackage{hyperref}% add hypertext capabilities
\usepackage{color}
\usepackage{tikz}
\usepackage{subcaption}
\begin{document}

\preprint{APS/123-QED}

\title{Symmetries and field tensor network states}
\author{Albert Gasull}  \email{albert.gasull@mpq.mpg.de}
 \affiliation{Max-Planck-Institut für Quantenoptik, Hans-Kopfermann-Strasse 1, D-85748 Garching, Germany}
 \author{Antoine Tilloy}
\affiliation{LPENS, Département de physique, École Normale Supérieure -- Centre Automatique et Systèmes (CAS), Mines Paris-- Université PSL, Sorbonne Université, CNRS, Inria, 75005 Paris}
\author{J. Ignacio Cirac$^{1,}$}
\affiliation{Munich Center for Quantum Science and Technology (MCQST), Schellingstrasse 4, D-80799 München, Germany}
\author{Germ{\'a}n Sierra}
\affiliation{
 Instituto de Física Teórica UAM/CSIC, Universidad Autónoma de Madrid, Cantoblanco, Madrid, Spain
}
\date{\today}

\begin{abstract}
We study the interplay between symmetry representations of the physical and virtual space on the class of tensor network states for critical spins systems known as field tensor network states (fTNS). These are by construction infinite dimensional tensor networks whose virtual space is described by a conformal field theory (CFT). We can represent a symmetry on the physical index as a commutator with the corresponding CFT current on the virtual space. By then studying this virtual space representation we can learn about the  critical symmetry protected topological properties of the state, akin to the classification of symmetry protected topological order for matrix product states. We use this to analytically derive the critical symmetry protected topological properties of the two ground states of the Majumdar-Ghosh point with respect to the previously defined symmetries.
\end{abstract}

\maketitle

%\tableofcontents

\section{Introduction}
Many-body systems are inherently hard to solve numerically due to the exponential size of their Hilbert space and analytical solutions are not always available. In such situations one usually relies in suitable ansatzes that capture the most relevant physics and symmetries of the system. One such class of states are Tensor Network States (TNS), which have been established as one of the most relevant tools to describe low-energy states of many non-trivial systems in different spatial dimensions \cite{Orusreview2014,TNreview2021}. In this description, physical degrees of freedom are associated to elementary tensors, connected into a network via auxiliary legs whose dimension (the bond dimension) bounds the amount of entanglement present in the system. TNS have several interesting built-in properties, such as the area law of entanglement \cite{GappedHandarealaw} and  they bring to the table a plethora of numerical methods, such as the celebrated DMRG \cite{OriginalDMRG}. On the analytical side, the interplay between the representation of symmetries in the virtual and physical degrees of freedom has also shed light into the classification of quantum phases of matter \cite{SymmetricTNS,WenScience?,SanzPRL,SanzPRA,VidalPRLtopo}. This classification was shown to fully characterize symmetry protected topological (SPT) order for the class of most basic one dimensional tensor network, Matrix Product States (MPS) \cite{SPTMPSclassification,MPSandSPT2011,GuWenMPS,PollmanBergMPS}, and to also characterize topological order in two dimensions for Projected Entangled Pair States (PEPS) \cite{PEPSandanyons,GuWenPEPS}. 

In recent years, there has been an extensive effort to apply TNS techniques to both Lattice Gauge Theories (LGT) \cite{LGTTNSreview} and Quantum Field Theories (QFT) \cite{cMPSVerstraete2015,cTNs2019,cMPS,cMERA,OsbornecPEPS}. While the first approach mostly relies on using numeric techniques from TNS to compute quantities in a discretized space, the former is based on extending the ansatz of TNS to the continuum limit. The first generalization of MPS in one dimension is known as a continuous MPS (cMPS) \cite{cMPSVerstraete2015} and has been applied to continuous bosonic systems in \cite{bosoniccMPS} and relativistic 1+1 dimensional quantum field theories in \cite{relativisticcMPS,OsbornecPEPS}. Generalizations to higher dimensions were brought forth in \cite{cTNs2019}, and were given the name of continuous TNS (cTNS), and have so far seen success in describing Gaussian QFT states \cite{GaussiancTNS}.

The fact that tensor network states like MPS obey the area law of entanglement by construction prevents them from exactly describing systems whose correlations are logarithmic in the size of the system, a situation that arises at critical points associated to phase transitions. As it was shown by Vidal et al. and Cardy in \cite{Criticalentanglement},\cite{CardyCalabrese2004}, the entanglement entropy for a critical one-dimensional system scales as $S_{\ell}\sim c\log{\ell}$ for a block of size $\ell$ in an infinite system, where $c$ is the central charge of the corresponding Conformal Field Theory (CFT). While it would be impossible for a standard MPS to exactly describe this situation, TNS that go beyond the area law and accurately describe such systems have been constructed, their prime example being the Multi Entanglement Renormalization Ansatz (MERA)\cite{MERAVidal}.

This motivates the introduction of another notion of continuum for TNS, where the physical space remains discrete (spins) but the auxiliary space becomes continuous. This corresponds to describing the state of the discrete spin system with long range interactions by means of a correlator of operators of an underlying CFT \cite{IMPS2010,IMPS2011}. This correlator can then be brought in the form of a path integral generalization of TNS, which was denoted field TNS (fTNS) in \cite{fTNS2021}. A natural question then arises, how much of what we know from the structure of discrete TNS, applies to their field theory equivalents? In particular, how are symmetries represented in the CFT virtual space connected to the physical ones?

In this paper we approach this question and seek to understand the SPT properties of the $\text{SU}(2)_1$ Wess-Zumino-Witten (WZW) fTNS, which include the critical one-dimensional ground states of the Haldane-Shastry model \cite{Shastry1988} by following a similar approach as it is done with MPS techniques. In MPS, one finds that the representation of the symmetry on the physical degrees of freedom can be mapped to the action of the symmetry on the corresponding discrete virtual space. The projective representation of the symmetry on the virtual space in turn classifies SPT order for one dimensional systems \cite{GuWenSPT}. We wish to establish an analogous statement for the case of fTNS, where the physical degree of freedom is discrete, and the virtual space corresponds to a CFT. In order to do so we establish a relation between the $\text{SU}(2)$ symmetry present in the physical space of the Haldane-Shastry model and the $\mathfrak{su}(2)_1$ conformal currents on the virtual space. We then find that the representation in the virtual space is in general projective and thus that fTNS can host critical SPT order.

This manuscript is organized as follows. In Section II we briefly review the fTNS construction. In section III we first briefly recap how the symmetries of discrete tensor networks provide insights on the topological properties of states, identify the relevant symmetries of the free boson fTNS and then show how we can relate the symmetries of the physical index to the virtual space. In section IV we then use this result to analyze the two possible ground states of the Majumdar-Ghosh model and find that we can distinguish them with our symmetry arguments.

\section{fTNS in one dimension}
We start by reviewing the construction of field tensor network states (fTNS) as presented in \cite{fTNS2021}. We emphasize throughout this section the connection of fTNS with standard TNS, as our results are naturally understood as analogous to those equivalent for TNS. We then provide the main expressions for the one-dimensional fTNS of the WZW SU(2)$_1$ fTNS.

We start by considering a system of $N$ d-dimensional spins and we write their wavefunction as 
\begin{equation}
    |\psi\rangle = \sum_{s_1...s_N=1}^dc_{s_1,...,s_N}|s_1 ... s_N\rangle.
    \label{wavefunction}
\end{equation}
A natural ansatz would be to represent the coefficients $c_{s_1,...,s_N}\in\mathbb{C}$ as a translationally invariant MPS
\begin{equation}
    \label{MPS}
    c_{s_1,...,s_N} = \sum^D_{n_1,...,n_N=1}A^{s_1}_{n_1,n_2}A^{s_2}_{n_2,n_3}...A^{s_N}_{n_N,n_1},
\end{equation}
where the matrices $A^{s_i}_{n_i,n_{i+1}}$ are a set of $d$ complex matrices of dimension $D\times D$ and the parameter $D$ is known as the bond dimension. fTNS can be intuitively understood as a generalization of MPS where the matrices are promoted to operators acting on 1 dimensional quantum field states as in equation (\ref{MPS}) 
\begin{equation}
    \label{FTNdef}
    c_{s_1,...,s_N}=\int\mathcal{D}[f_1]...\mathcal{D}[f_N]\mathcal{A}_{f_1,f_2}^{s_1}...\mathcal{A}_{f_N,f_1}^{s_N}.
\end{equation}
This replaces the previously discrete indices of the matrices for functions $f_i : \mathbb{R}\rightarrow\mathbb{R} \in \mathbb{L}^2(\mathbb{R})\cup\mathbb{K}$, where $\mathbb{K}$ denotes the set of constant functions and the sum over indices  gets replaced by a path integral. Thus, for every value of the spin $s_i$, $\mathcal{A}_{f_i,f_{i+1}}^{s_i}$ is a functional of both $f_i$ and $f_{i+1}$ making this a generalization that takes finite matrices to functionals and discrete sums to path integrals over the space of square integrable functions.

\begin{figure}[h!]
\begin{subfigure}{\linewidth}
\centering
\subcaption{ }
\scalebox{0.8}{\includegraphics[width=\linewidth]{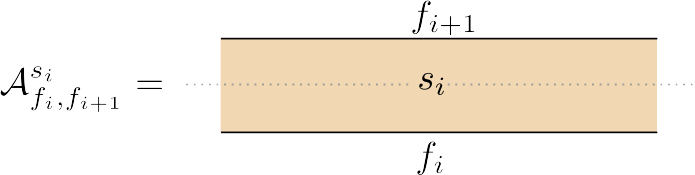}}
\end{subfigure}
\begin{subfigure}{\linewidth}
\centering
\caption{ }
\includegraphics[width=\linewidth]{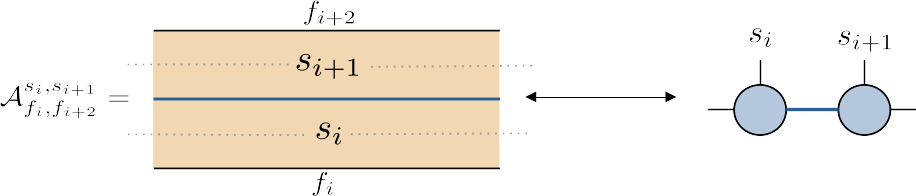}
\end{subfigure}
\begin{subfigure}{\linewidth}
\centering
\caption{ }
\includegraphics[width=\linewidth]{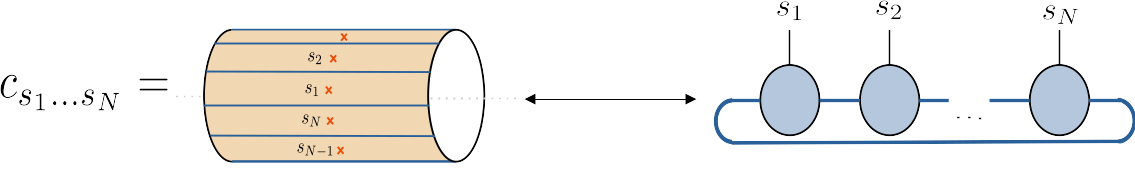}
\end{subfigure}
\captionsetup{justification=centerfirst}
\caption{a) Diagrammatic representation of the functional $\mathcal{A}^{s_i}_{f_i,f_{i+1}}$. The upper and lower boundaries of the strip are the support of the boundary functions which serve as indices for the functional. b)-c) Diagrammatic representation of the sewing of two strips and the closing condition, alongside their MPS equivalent operation.}
\label{diagrammatics}
\end{figure}

Similarly to the contraction of the virtual indices in standard TNS theory, we define the "sewing condition" by
\begin{equation}
   \label{sewing}
   \mathcal{A}^{s_1...s_m}_{f_1,...,f_{m+1}}=\int\mathcal{D}[f_2]...\mathcal{D}[f_m]\mathcal{A}^{s_1}_{f_1,f_2}...\mathcal{A}^{s_m}_{f_m,f_{m+1}},
\end{equation}
for $m<N$, which consists of the contraction of $m$ functionals out of the $N$ total that make the complete wavefunction. To recover the complete wavefunction we hence need to multiply all $N$ functionals, and for periodic boundary conditions, integrate over the remaining free indices
\begin{equation}
    \label{closing}
    c_{s_1,...s_N} = \int\mathcal{D}f\mathcal{A}^{s_1,...,s_N}_{f,f},
\end{equation}
which we call the ``closing condition". We can illustrate all of these operations diagrammatically with strips as in figure (\ref{diagrammatics}).

Of particular interest are those wavefunctions whose coefficients can be written in terms of correlators of a 2D CFT with a local action, since its inherent power law correlations prevents an MPS description with finite bond dimension \cite{IMPS2010}. In this work we focus our attention on the free massless boson, which is one of the simplest CFTs  and can be also understood as the WZW SU(2)$_1$ CFT. The coefficients for this state can be written as
\begin{equation}
    \label{coeffs}
    c_{s_1,...s_N}\propto\langle\chi_{s_1}:e^{is_1\sqrt{\alpha}\phi(z_1)}:...\chi_{s_N}:e^{is_N\sqrt{\alpha}\phi(z_N)}: \rangle_0,
\end{equation}
where $\chi_{s_i}$ is a phase factor that can depend on $s_i$, :: denotes normal ordering, $\phi(z_i)$ is the chiral real massless scalar field and the subscript 0 denotes the correlator taken in the vacuum of the CFT. The chiral vertex operators $:e^{is_i\sqrt{\alpha}\phi(z_i)}:$ with $\alpha=\frac{1}{2}$ are the spin $\frac{1}{2}$ primary fields of the WZW SU(2)$_1$ theory. This family of coefficients includes the groundstate of the Haldane-Shastry model \cite{Haldane1,Haldane2} which is defined by the long range Hamiltonian
\begin{equation}
    \mathcal{H}_{HS}=-\sum_{i\neq j }\frac{z_iz_j}{\left(z_i-z_j\right)^2}\left(\mathcal{P}_{ij}-1\right),
\end{equation}
where $z_i$ are the positions of the spin and $\mathcal{P}_{ij}$ is the spin permutation operator. This is a paradigmatic model of criticality, and its groundstate can be obtained from  the state defined in equation (\ref{coeffs}) by choosing $s_i=\pm1$, $\alpha=\frac{1}{2}$,  $\chi_{s_m}=e^{im\pi(s_m-1)/2}$ and defining the CFT to be on a cylinder of circumference $\pi N$ \cite{Yellowbook}. The latter choice yields the following power-law-like coefficients
\begin{equation}
    \label{HScoeffs}
    c_{s_1,...s_N}\propto\delta_{\sum_ns_n,0}\prod_n\chi_{s_n}\prod_{n>m}\left(\sin\left[\frac{(n-m)}{N}\right]\right)^{\frac{s_ns_m}{4}},
\end{equation}
when the spins are located in the cylinder at positions $z_n =e^{\frac{2\pi i n}{N}} $ which correspond to the positions from equation (\ref{coeffs}) of the vertex operator insertions. We quickly outline how to construct the corresponding fTNS for CFTs with a local solvable action with the aforementioned free boson.

\begin{figure}[h!]
\includegraphics[width=\linewidth]{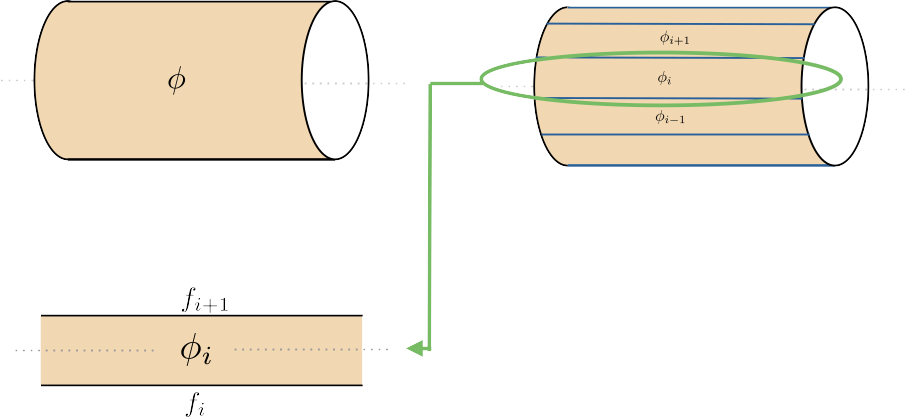}
\captionsetup{justification=centerlast}
\caption{\label{PIsplitfig} Visual representation of the splitting of the functional integral of equation (\ref{PIsplit}).}
\end{figure}

The first step in identifying the corresponding fTNS is to write the correlator of equation (\ref{coeffs}) in the functional integral representation as
\begin{equation}
    \label{PIcoeffs}
    c_{s_1,...,s_N}\propto\int\mathcal{D}[\phi]e^{-S_E}e^{-i\sum_{n=1}^{N}s_n\sqrt{\alpha}\phi(\textbf{z}_n)}\prod_n\chi_{s_n},
\end{equation}
where $S_E$ is the euclidean action of the free massless boson $\phi : \mathbb{R}\rightarrow\mathbb{R}$ defined on the cylinder of length $\pi N$. We can now split this functional integral in different regions as
\begin{equation}
    \label{PIsplit}
    \int\mathcal{D}[\phi]=\int\mathcal{D}[f_1]...\mathcal{D}[f_N]\int^{'}\mathcal{D}[\phi_1]...\int^{'}\mathcal{D}[\phi_N],
\end{equation}
where the functions $f_i$ corresponds to the boundary conditions between neighbouring regions. After the splitting, the scalar fields $\phi_{i}$ are defined only within their respective strips as shown in figure \ref{PIsplitfig}. Note here, that we place the cuts so that each of the resulting strips encloses only one of the vertex operators of (\ref{coeffs}).

The integrals with a prime must respect the fields boundary conditions given by
\begin{equation}
\begin{split}
    &\phi(z_i^+)=f_{i+1}(x) \\ 
    &\phi(z_i^-)=f_{i}(x),
\end{split}
\end{equation}
where $z_i^{\pm}$ are the corresponding positions of the boundaries of the $i^{\text{th}}$ strip and $x$ is the coordinate along the boundary.

Finally one identifies the functional $\mathcal{A}^{s_i}_{f_i,f_{i+1}}$ with the value of the $i^{\text{th}}$ strip given by
\begin{equation}
    \label{functionaldef}
    \mathcal{A}^{s_i}_{f_i,f_{i+1}} = \int^{'}\mathcal{D}[\phi_i]e^{-S}e^{-is_i\sqrt{\alpha}\phi_i(z_i)}\chi_{s_i}.
\end{equation}
In \cite{fTNS2021}, the complete derivation of this functional was done by means of Green function techniques, and their sewing and closing conditions were computed. To recover the coefficients of equation (\ref{HScoeffs}) with fTNS it is necessary to truncate the functional only to its chiral component, similar to how the correlator in equation (\ref{coeffs}) is computed using only the chiral sector of the corresponding CFT. At the same time, in order to obtain the condition $\delta_{\sum_{n=1}^{N}s_n,0}$ that arises from the zero mode of the boson field, the set of square integrable functions alone is not enough, which is why the set of constant functions was needed for the set of possible strip boundary functions.

The free boson  fTNS was studied in depth in \cite{fTNS2021}, and we recover here the explicit expressions for clarity. Each of the functionals is defined on a strip $\mathcal{M}_{\Delta_j} = \mathbb{R}\times i \pi \left[a_j,b_j\right]$ with width $\Delta_j = b_j-a_j$. We denote the functionals $\mathcal{A}_{f_i,f_{i+1}}^{s_i} = \mathcal{A}_{\Delta_i}\left[f_0,f_i,f_{i+1},\{z_i,s_i\}\right]$, as we wish to make more explicit the dependence of the functionals on the boundary functions and physical degrees of freedom, where $z_i\in\mathcal{M}_{\Delta_i}$ is the position of the $i^{\text{th}}$ spin and $f_0$ is the 0-mode which is assumed to be the same for all functionals. The expression for a single functional is then given by 
 \begin{equation}
 \label{amplitude}
\begin{split}
    & \mathcal{A}_{{\Delta_i}}\left[f_0,f_+,f_-,\left\{z_i,s_i\right\}\right] = e^{f_0\frac{s_i}{\sqrt{2}}}e^{-R_{{\Delta_i}}\left[f_+,f_-,\left\{z_i,s_i\right\}\right]}, \\
    & R_{{\Delta_i}}\left[f_+,f_-,\left\{z_i,s_i\right\}\right]= \frac{s_i^2}{4}\log{\Delta_i} \\
    &+\frac{1}{2}\int^{\infty}_{0}\text{d}k
\begin{pmatrix}
\hat{f}_{+}(k) & \hat{f}_{-}(k) 
\end{pmatrix}
\begin{pmatrix}
\omega_{+,\Delta_i}(k) & \omega_{-,\Delta_i}(k) \\
\omega_{-,\Delta_i}(k) & \omega_{+,\Delta_i}(k)
\end{pmatrix}
\begin{pmatrix}
\hat{f}_{+}^{*}(k) \\
\hat{f}_{-}^{*}(k)
\end{pmatrix} \\
&-\frac{i}{2\sqrt{2}}s_i\int_{\mathbb{R}}\text{d}k\frac{e^{ikz_i}}{\sinh{(\pi k \Delta_i)}}\left(e^{\pi k b_i }\hat{f}_{+}(k)-e^{\pi k a_i}\hat{f}_{-}(k)\right),
\end{split}
\end{equation}
with $\omega_{+,\Delta_i} = k\coth{(\pi k \Delta_i)}$ and $\omega_{-,\Delta_i} = -k \ \text{sech}{(\pi k \Delta_i)}$ and $\hat{f}_{\pm}(k):\mathbb{R}\rightarrow\mathbb{R}$ the Fourier transform of the corresponding lower and upper boundary functions $f_{\pm}(x)$. With this expressions the sewing condition reads
\begin{equation}
\label{ftnsewing}
\begin{split}
     & \int[\text{d}g]\mathcal{A}_{{\Delta_1}}\left[f_0,f_+,g,\left\{z_1,s_1\right\}\right]\mathcal{A}_{{\Delta_2}}\left[f_0,g,f_-,\left\{z_2,s_2\right\}\right]  \\ &= \mathcal{A}_{{\Delta_1\cup\Delta_2}}\left[f_0,f_+,f_-,\left\{z_i,s_i\right\}_{i=1,2}\right],
    \end{split}
\end{equation}
 where now the exponent of the sewn strips is given by :
\begin{equation}
\label{ftnsewingbig}
    \begin{split}
    & R_{{\Delta_1\cup\Delta_2}}\left[f_+,f_-,\left\{z_i,s_i\right\}_{i=1,2}\right]  = \frac{s_1^2}{4}\log{\Delta_f}+\frac{s_2^2}{4}\log{\Delta_f}\\
    &+\frac{1}{2}\int^{\infty}_{0}\text{d}k
\begin{pmatrix}
\hat{f}_{+}(k) & \hat{f}_{-}(k) 
\end{pmatrix}
\begin{pmatrix}
\omega_{+,\Delta_f}(k) & \omega_{-,\Delta_f}(k) \\
\omega_{-,\Delta_f}(k) & \omega_{+,\Delta_f}(k)
\end{pmatrix}
\begin{pmatrix}
\hat{f}_{+}^{*}(k) \\
\hat{f}_{-}^{*}(k)
\end{pmatrix} \\
&-\frac{i}{2\sqrt{2}}\int_{\mathbb{R}}\text{d}k\frac{\sum_{i=1,2}e^{ikz_i}s_i}{\sinh{(\pi k \Delta_f)}}\left(e^{\pi k b_2 }\hat{f}_{+}(k)-e^{\pi k a_1}\hat{f}_{-}(k)\right) \\ 
& - \frac{s_1s_2}{2}\log{\left(\mu\sinh{\left(\frac{z_2-z_1}{2
\Delta_f}\right)}\right)},
\end{split}
\end{equation}
where $\Delta_f=\Delta_1+\Delta_2$ and $\mu=-2i$. These are the main expressions that will be used in the upcoming sections to study the symmetry properties of the states defined by this functional.

\section{Symmetry relations for MPS and fTNS}
We now turn our attention to the study of the symmetries of the functional $\mathcal{A}_{{\Delta}}[f_0,f_+,f_-,\{z_i,s_i\}]$, where from now on the $0$-mode $f_0$ will be omitted as it does not participate in the computation of the properties that we wish to explore. We draw our intuition from the results of standard MPS theory, and thus briefly recall them now. When a state is symmetric under a representation $U_g$ of a symmetry group $G$, and can be represented by an injective MPS in canonical form \cite{TNreview2021}, then the following relation holds
\begin{equation}
\label{MPSsymmetry}
\includegraphics[width=0.75\linewidth]{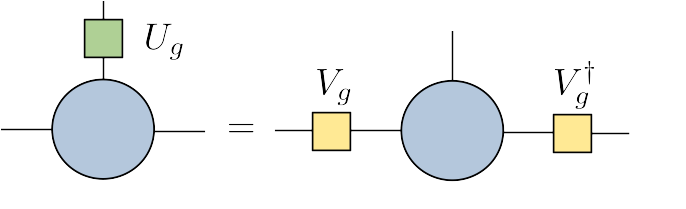}
\end{equation}
where the representation on the virtual space $V_g$ can in general be a projective representation \cite{SanzPRA,PollmanBergMPS}. In equation form, 
\begin{equation}
    \sum_{s_j}(U_g)_{s_i,s_j}A^{s_j}_{n,m}=\sum_{k,l}(V_g)_{n,k}A^{s_j}_{k,l}(V_g^{\dagger})_{l,m},
\end{equation}
 which holds for instance when $G$ is a Lie group and we can use the exponential map to write $U_g(\theta)=e^{i\theta\mathfrak{g}}$, where $\mathfrak{g}$ is an element of the corresponding Lie algebra. Projective representations differ from linear ones in that they fulfill the more general composition rule
\begin{equation}
\label{projectivity}
    V_gV_h=e^{i\omega(g,h)}V_{gh},
\end{equation}
where the extra phase factor is known as the cocycle $\omega(g,h)\in \mathcal{H}^2(G,\text{U}(1))$ which in turn is known to fully classify SPT order in one dimension \cite{TNreview2021,GuWenMPS,PollmanBergMPS,MPSandSPT2011}. Equation (\ref{projectivity}) is the main relation that we wish to establish in the context of fTNS. We first start by identifying the relevant symmetries and how they are represented on both the discrete physical index and the continuous functional space. 

The physical symmetry of interest is the SU(2) symmetry that is present in the Haldane-Shastry model, whose critical point is accurately described by the massless free boson \cite{ZaletelHaldaneCFT}. On the one hand, the symmetry on the physical indices is simply the one corresponding to a discrete two-dimensional spin and it is thus represented by the usual Pauli matrices that generate SU(2). On the other hand, the virtual space of the fTNS is given by a CFT, specifically the SU(2)$_1$ WZW model. For this CFT, the symmetry is represented by the conformal charges of the currents corresponding to the affine simple Lie algebra of the model. As presented in \cite{Yellowbook}, the conformal currents that generate $\hat{\mathfrak{su}}(2)_1$ are given in terms of fields as
\begin{equation}
    \label{affineliealgebra}
\begin{split}
     & \text{H}(z) = :i\partial\phi(z):, \\
     & \text{E}^{\pm}(z) = :e^{\pm i\sqrt{2}\phi(z)}:,
\end{split}
\end{equation}
 where $\phi(z)$ is the chiral part of the free massless boson, :: denotes normal ordering and we can think of H and E$^{\pm}$ as analogous to $\sigma^z$ and $\sigma^{\pm}$ respectively of the more familiar $\mathfrak{su}(2)$ algebra.
 
In order to represent these currents as functionals, we need to recall that the primary fields of the theory of the massless boson are given, in the chiral sector, by the vertex operators $:e^{\pm i\frac{1}{\sqrt{2}}\phi(z)}:$ with conformal dimension $h=\frac{1}{4}$. We can thus represent each of them by the corresponding fTNS functionals $\mathcal{A}_{{\Delta_i}}[f_+,f_-,\{z_i,\pm\frac{1}{\sqrt{2}}\}]$. Thanks to the state-operator correspondence, we can think of these operators as the ones generating the states corresponding to the eigenvectors of $\sigma_z$ in the usual 2-dimensional spin picture \cite{Yellowbook}. That is, the OPE of $:e^{\pm i\frac{1}{\sqrt{2}}\phi(z)}:$ with $E^{\pm}(w)$ has no singular terms in $z-w$, which results in no contribution upon computing the contour integral corresponding to the commutator in the CFT picture. This means that in the usual spin picture, $E^{\pm}(z)$ is the equivalent  of the ladder operator $\sigma_{\pm}$.  Similarly to how we have represented a vertex operator as a functional $\mathcal{A}_{{\Delta_i}}$, we must also write the currents as functionals as well.
The $\text{U}(1)$ current of the algebra is defined by the functional
\begin{equation}
\label{u1functional}
    H(z)_{{\Delta}} = \sqrt{2}\lim_{q\rightarrow 0} \frac{1}{q}\partial_z \mathcal{A}_{{\Delta}}\left[f_+,f_-,\left\{z,q\right\}\right],
\end{equation}
while the SU(2) currents are
\begin{equation}
    \label{su2functional}
    E^{\pm}(z)_{{\Delta}} = -\frac{\mu}{2}{A}_{{\Delta}}[f_0,f_+,f_-,\left\{z,\pm\sqrt{2}\right\}].
\end{equation}

It is important to notice that since all conformal currents have conformal dimension $h=1$, the limit of $\Delta\rightarrow0$ must be taken, but this limit can only be taken after they have been applied to another functional representing a state. This is testament to the fact that these operators are in fact defined as distributions and hence they must be understood in terms of how they act on a set of test functionals, in our case the vertex operator functionals. To more physically understand this limit, let us think of the effect of adding a symmetry charge inside of a correlator of total length $L$. Upon sewing and closing all the strips, we would end up with a system that has length $L+\Delta$, and thus to recover back the original physical length of the system we need to take the limit of the width of operators to zero, that is $\Delta\rightarrow 0$ in both equation (\ref{u1functional}) and (\ref{su2functional}).

 At this point, the representation of the symmetry in terms of conformal currents $J(z)$ is not yet complete, as the corresponding symmetry operators on the virtual space of the fTNS must be identified with the conformal charges given by $Q=\frac{1}{2\pi i}\oint\text{d}zJ(z)$, where the contour encircles the origin of the complex plane. However, this identification is not yet completely correct as we must take into account that the functional $\mathcal{A}_{\Delta}$ is defined on a strip, and not the whole complex plane as in usual CFT. Such a manifold with boundaries can only be mapped to the upper half part of the complex plane (UHP) because of the Riemann mapping theorem (see the supplementary material of \cite{fTNS2021}). This means that we must define the charges to be contained in the UHP as well by means of the method of images, as it is done in the setting of boundary CFT \cite{Yellowbook}, which reads
 \begin{equation}
     \label{chargedef}
     Q = \frac{1}{2\pi i }\oint \text{d}zJ(z) =\frac{1}{2\pi i}\int_{\includegraphics[scale=0.25]{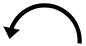}}\text{d}z\left(J(z)-J(\bar{z})\right).
 \end{equation}

As is usual in Lie group theory, parametrizing the unitaries in equation (\ref{MPSsymmetry}) in terms of a small angle $\theta$ and differentiating on both sides, one can obtain the equivalent relation for the algebra, which for fTNS reads
\begin{equation}
\label{algebrarelation}
\includegraphics[width=0.9\linewidth]{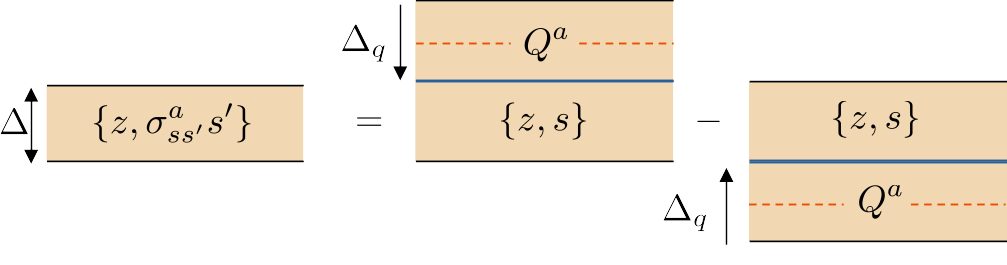},
\end{equation}
where $\Delta_q$ is the width associated to the symmetry functional which is then taken $\Delta_q\rightarrow 0$ and the red line indicates the current integration. We want to emphasize that on the l.h.s of equation (\ref{algebrarelation}) the symmetry is represented by a finite $2\times2$ matrix acting on the physical spin, which is denoted by $\sigma^a s$, while on the r.h.s it is represented by the commutation with a conformal charge, which is essentially an infinite dimensional functional. It is then shown in the appendix that the result of acting with the physical algebra via their action through the virtual space, that is equation (\ref{algebrarelation}), is given by

\begin{align}
    \label{algebraaction}
    &\sigma^z_{ss'}\mathcal{A}_{{\Delta}}\left[z,\frac{s'}{\sqrt{2}}\right]= s \mathcal{A}_{{\Delta}}\left[z,\frac{s}{\sqrt{2}}\right], \\
    &
    \sigma^{\pm}_{ss'}\mathcal{A}_{{\Delta}}\left[z,\frac{s'}{\sqrt{2}}\right] = \begin{cases}
    & 0 , \ \ (s=\pm1) \\
    &\mathcal{A}_{{\Delta}}\left[z,\frac{s}{\sqrt{2}}\pm\sqrt{2}\right] , \ (s=\mp1), \notag\\
\end{cases}
\end{align}
where the functionals with $s=\pm1$ are the two primary fields of the model which are equivalent to the two possible spin projections and we have dropped the argument of the boundary functions for simplicity. This shows how the action of the symmetry on the virtual space is equivalent to the one we would expect on the physical index.
These results agree with what one would get from the same computation by means of  the CFT formalism by computing the operator product expansion (OPE) between the different currents and states. It is worth noting that it is possible to identify equation (\ref{algebrarelation}) with the contour integral present in most of the aforementioned CFT computations.

Not only can we compute equation (\ref{algebrarelation}) for all elements of the algebra, but also each of the individual terms  of this equation separately, which is very important in order to be able to compute the algebra relations, and hence the cocycle. The complete set of relations (see  equations (\ref{begin})-(\ref{end}) in the appendix) is given by :
\begin{equation}
    \label{sigmax}
    \includegraphics[width=\linewidth]{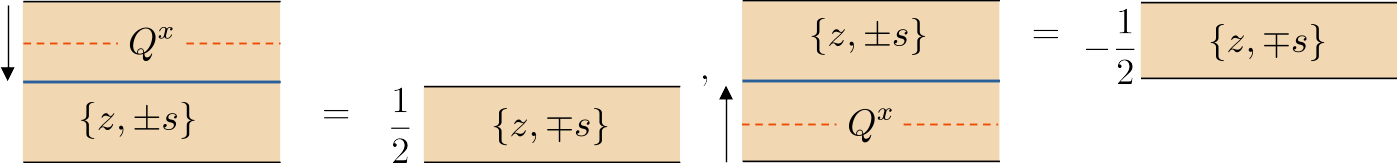},
\end{equation}
\begin{equation}
    \label{sigmay}
    \includegraphics[width=\linewidth]{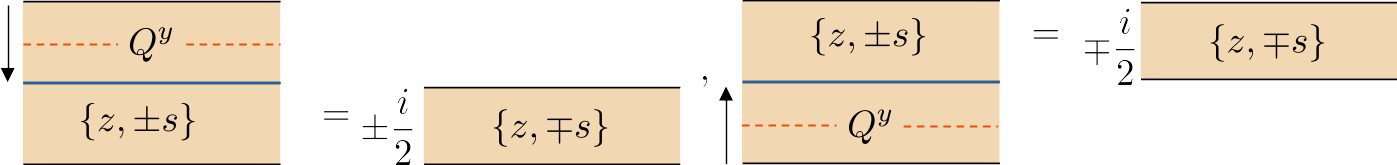},
\end{equation}
\begin{equation}
    \label{sigmaz}
    \includegraphics[width=\linewidth]{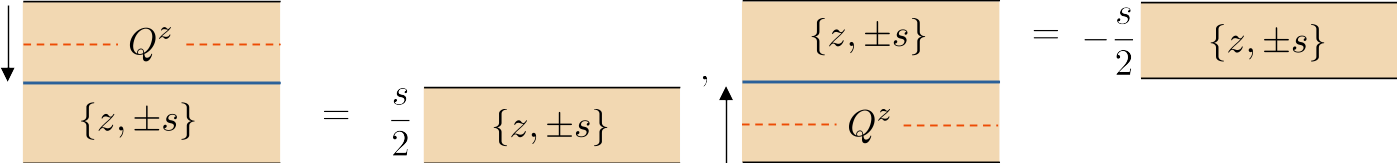},
\end{equation}
where the r.h.s strip is always aligned with the physical strip of the l.h.s onto which the charge is applied, as otherwise we would be comparing strips on different coordinate systems. As shown in the appendix, we have taken the upper and lower limits of the strip to be at $\pm\frac{\pi\Delta}{2}$ and the functional representing the action of $\sigma^x$ is computed from $\sigma^x=\frac{1}{2}(\sigma^{+}+\sigma^{-})$ and similarly for $\sigma^y$.  Due to the limit of the charge strip width going to zero, to compute the algebraic relations between the different charges we must do so by applying all of them onto a test functional. With this set of rules we can then compute the group commutator to extract the cocycle, which we can write as 
\begin{equation}
    \label{groupcommstate}
    \includegraphics[width=\linewidth]{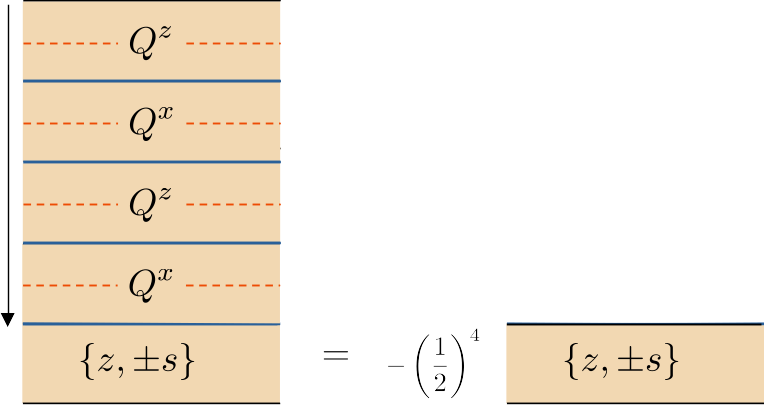},
\end{equation}
where the negative sign in front confirms that we have a projective representation. It is important to point out that in equation (\ref{groupcommstate}), or whenever we encounter terms that consist of more than two strips being sewn, one must be consistent in the order in which the sewings take place. As shown in the appendix, we have used a bottom-to-top order and, as long as one remains consistent, the end result is independent of the specific order chosen. With all these relations, we can also prove the SU(2) symmetry of the original state in this picture, as it is shown in the appendix.

To summarize, we have represented the relevant conformal charges as functional strips, computed their action on their corresponding fTNS, and furthermore, evaluated their group commutator to conclude that they indeed form a projective representation. With this toolbox in our hand we can now study how the positions $z_i$ of the different spin insertions affects the symmetries of the functionals and hence the properties of the state.

\section{Distinct SPT ground states of the Majumdar-Ghosh point}

Ansaetze based upon vertex operators were used in \cite{IMPS2010} to study the properties of several models whose critical points were accurately described by a $c=1$ CFT. The positions of the vertex operator insertions were treated as the variational parameters in order to maximize the overlap with the real groundstate. Can we use the positions of the spins in fTNS to change the physical properties of the state they describe? We investigate this question by studying different limits for the spin insertions and take a look at the SU(2) representation of the virtual space to identify the corresponding phase of the system.

\begin{figure}[h!]
    \centering
    \includegraphics[width=\linewidth]{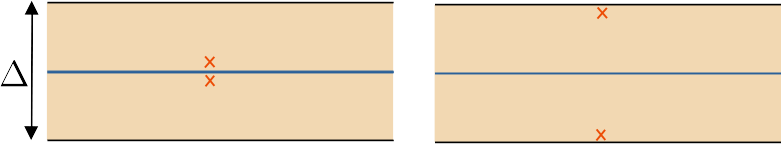}
    \captionsetup{justification=centerlast}
    \caption{This figure shows the two limits of interest for a pair of spin insertions. The left strip corresponds to the trivial representation while the right strip belongs to the non-trivial one.}
    \label{dimer}
\end{figure}

First we consider the limit in which 2 spins are placed very close together as shown in the left  figure (\ref{dimer}), which in CFT literature is the limit that one must consider to compute OPEs \cite{Yellowbook}. We know  from the CFT literature \cite{WZW2011,Yellowbook} that for the WZW SU(2)$_k$ model, the fusion rules for the $k+1$ primary fields $\phi_j$ with $2j+1$ components are
\begin{equation}
    \label{fusionrules}
    \phi_{j_1}\times\phi_{j_2}=\sum^{\text{min}(j_1+j_2,k-j_1-j_2)}_{j=|j_1-j_2|}\phi_j,
\end{equation}
which for SU(2)$_1$ reduce to $\phi_{\frac{1}{2}}\times\phi_{\frac{1}{2}}=\phi_0$. This means, that whenever two spins are very close, CFT tells us that the dominant term in the expansion should be the identity. By taking the limit $z_2\rightarrow z_1$ in equation (\ref{ftnsewingbig}), similar to the one taken when performing an OPE, the expression for a strip with two spins in this limit becomes
\begin{align}
    \label{ftnsinglet}
     &\lim_{z_1\rightarrow z_2} \mathcal{A}_{{\Delta}}[f_+,f_-,\{z_1,s_1,z_2,s_2\}]\sim \\  &\frac{\mu^{\frac{s_1s_2}{2}}\delta_{s1,-s2}\sqrt{2\Delta}}{\sqrt{z_1-z_2}}\mathcal{A}_{{\Delta}}[f_+,f_-,\{z_1,0\}]= \notag \\ &\frac{\sqrt{2}\delta_{s1,-s2}}{\mu\sqrt{z_1-z_2}}\mathbb{I}_{{\Delta}}[f_+,f_-] \notag,
\end{align}
where $\delta_{s1,-s2}$ ensures the spins have opposite value and $\sim$ means that we have omitted sub-leading terms in $z_1-z_2$. Remarkably, whenever two insertions get close to each other, the functional greatly simplifies and becomes an identity in the virtual space. The decoupling  of the virtual space from the physical space is a phenomenon one encounters as well when considering dimerized states in MPS theory \cite{SPTMPSclassification}. To mimic the results of MPS, we are interested in seeing how the symmetry is represented in this limit . We can see by applying the rules (\ref{sigmax}),(\ref{sigmay}),(\ref{sigmaz}) on the identity, that since it corresponds to a strip with $s=0$, the outcome is always 0. This is akin to how the monomial representation of $\mathfrak{su}(2)$ acts on the $j=0$ element. The main point to take away, is that this limit forces the virtual space to be on the trivial representation $j=0$ of SU(2), making all the symmetry operators be simply the identity.

To obtain the $j=\frac{1}{2}$ we consider the opposite limit, in which two spins are placed as far apart from one another as possible as shown in the right figure of \ref{dimer}. Because of the inherent long-range interaction of the strips, we can only clearly understand the virtual space representation of any of the two boundaries when we take the limit $\Delta\rightarrow\infty$. Let us first study this limit for a single spin, in which we can approach the limit in different ways. We could send the upper (lower) boundary of the strip to the $(-)\infty$ limit, or both of them at the same time. In either case, whenever the spin is not located exactly at the boundary, the functional simplifies to
\begin{equation}
\lim_{\Delta\rightarrow\infty}\mathcal{A}_{\Delta}[f_+,f_-,\{z,s\}] = \frac{1}{\Delta^{\frac{s^2}{4}}}\mathbb{I}_{\infty}[f_+,f_-]
\label{equation31},
\end{equation}
where $\mathbb{I}_{\infty}[f_+,f_-]$ stands for the corresponding identity on virtual space for an infinitely wide strip. However, whenever the spin is sitting exactly at one of the boundaries being taken to infinity, the virtual space does not fully trivialize and instead remains within the corresponding spin representation, and we denote this limit by
\begin{equation}
    \lim_{\Delta\rightarrow\infty}\mathcal{A}_{\Delta}[f_+,f_-,\{z_b,s\}] = \frac{1}{\Delta^{\frac{s^2}{4}}}\mathcal{A}_{\infty}[f_+,f_-,\{z_b,s\}],
    \label{equation32}
\end{equation}
where $z_b$ is $i\pi b \  (i\pi a)$ for the upper (lower) boundary. The explicit expression corresponding to equation (\ref{equation31}) is
 \begin{equation}
 \label{infinityidentity}
\begin{split}
    & \mathbb{I}_{{\infty}}\left[f_+,f_-\right] = e^{-R_{{\infty}}\left[f_+,f_-\right]}, \\
    & R_{\infty}\left[f_+,f_-\right]  =  \\
    &+\frac{1}{2}\int^{\infty}_{0}\text{d}k
\begin{pmatrix}
\hat{f}_{+}(k) & \hat{f}_{-}(k) 
\end{pmatrix}
\begin{pmatrix}
k & 0 \\
0 & k
\end{pmatrix}
\begin{pmatrix}
\hat{f}_{+}^{*}(k) \\
\hat{f}_{-}^{*}(k)
\end{pmatrix}
,
\end{split}
\end{equation}
and the one for equation (\ref{equation32}) is
 \begin{equation}
 \label{infinityspin}
\begin{split}
    & \mathcal{A}_{{\infty}}\left[f_+,f_-,\{z_b,s\}\right] = e^{-R_{{\infty}}\left[f_+,f_-,\{z_b,s\}\right]}, \\
    & R_{\infty}\left[f_+,f_-,\left\{z_b,s\right\}\right]  =  \\
    &+\frac{1}{2}\int^{\infty}_{0}\text{d}k
\begin{pmatrix}
\hat{f}_{+}(k) & \hat{f}_{-}(k) 
\end{pmatrix}
\begin{pmatrix}
k & 0 \\
0 & k
\end{pmatrix}
\begin{pmatrix}
\hat{f}_{+}^{*}(k) \\
\hat{f}_{-}^{*}(k)
\end{pmatrix}\\
&-\frac{i}{2\sqrt{2}}s\int_{\mathbb{R}}\text{d}k\hat{f}_b(k),
\end{split}
\end{equation}
where the contribution of the zero mode has been omitted and $\hat{f}_b(k)$ is the corresponding boundary function of whichever boundary the spin is located at. We can then apply the rules (\ref{sigmax}),(\ref{sigmay}),(\ref{sigmaz}), which were derived in a $\Delta$-independent fashion, to conclude that the boundary of the strip at which the spin sits remains in the $s=\frac{1}{2}$. Similar expressions are obtained whenever we have several spins within the strip, the only contributions surviving the infinite width limit being the boundary ones. We can hence see, that in this limit the virtual space representation is completely dominated by whichever spin is located exactly at the boundary, and is hence non-trivial.

When we consider the case of finite $\Delta$, we can detect when one representation is favored by parameterizing the spin insertions by their distance away from the translation symmetric configuration. Let us take the case of 2 insertions, whose positions are parameterized by $z_1=i\pi a -\frac{i\pi\Delta}{4}\mp i\pi\delta$ and $z_2=i\pi a +\frac{i\pi\Delta}{4}\pm i\pi\delta$, where the term $i\pi a$ is there to ensure our choice of coordinate axis for the insertions does not matter. With these explicit positions, the 2-spin functional reads
\begin{equation}
\label{phasetransition}
    \left(\frac{i\mu}{\sqrt{2}}\right)^{\frac{s_1s_2}{2}}\left(\cos\left(\frac{\delta\pi}{\Delta}\right)\pm \sin\left(\frac{\delta\pi}{\Delta}\right)\right)^{\frac{s_1s_2}{2}}\mathcal{A}^{*}_{\Delta},
\end{equation}
where by $\mathcal{A}_{\Delta}^{*}$ we mean the 2 strip functional without the interaction term between the spins, that is without the last line of equation (\ref{ftnsewing}). We can then check which values of $\delta$ maximize this expression and how these relate to the different phases. We can see that whenever  the spins have opposite value $\frac{s_1s_2}{2}=-1$, the expression (\ref{phasetransition}) diverges for $\delta=\pm\frac{\Delta}{4}$, which exactly corresponds to the configuration presented in equation (\ref{ftnsinglet}). These positions corresponds to the spins meeting at the center of the system, and as we have seen this situation corresponds to the virtual space trivializing. Once we close the strip, the charge-neutrality condition prevents the strip with $\frac{s_1s_2}{2}=+1$ from contributing, however, we can still see which representations are favored in this case. We find that the maximum happens as well for $\delta=\pm\frac{\Delta}{4}$, in which case the functional simply inherits the representation of the spin closest to each boundary, as that is the dominant term as we take the $\Delta\rightarrow\infty$ limit. We thus see, that as soon as the insertions depart from the perfect spacing, immediately one of the two representations becomes favored depending on which pairing is encouraged.

We can now take the spin configuration on the left of figure \ref{dimer} a step further for the case in which we have more than two spins. Let us start with 4 spins and consider the limit $z_1\rightarrow z_2$ and $z_3\rightarrow z_4$, which corresponds to a situation like in equation (\ref{ftnsinglet}). If the distance between any two spins is denoted by $z_i-z_j=z_{ij}$, in the limit where $z_{12},z_{34}\rightarrow 0$ the four-spin functional becomes
\begin{align}
    &\frac{2\delta_{s_1,-s_2}\delta_{s_3,-s_4}}{\mu\sqrt{z_{12}}\sqrt{z_{34}}}\mathbb{I}_{\Delta}[f_+,f_-]+ \\ &\frac{\delta_{s_1,s_2}\delta_{s_3,s_4}\delta_{s_1,-s_3}\mu\sqrt{z_{12}}\sqrt{z_{34}}}{2}\mathcal{A}_{\Delta}[f_+,f_-,\{z_i^{*},2s_i^*\}_{i=1}^2]\notag,    
\end{align}
where $z_i^*$ are the positions at which the different pair of spins meet, and $s_i^*$ is the value of any of the two original spins of the pair. We see that the dominant term is the expected identity on the virtual space as it arises from the charge neutrality condition. However, having two pairs allows for the individual pairs to not have opposite spin value, but for the different pairs to compensate each others sign, and thus a new sub-leading term can arise. This sub-leading term corresponds exactly to a strip containing 2 spins of higher value, and thus a state constructed out of this term falls into a higher $SU(2)$ representation from the original one.

The different limits explored in this section are useful as they also correspond to the two distinct topological ground states of the Majumdar-Ghosh point of the $J_1$-$J_2$ Heisenberg model on $N$ sites, defined by 
\begin{equation}
\mathcal{H}_{J_1,J_2}=\sum_{i=1}^{N}\left(J_1\vec{S}_i\cdot\vec{S}_{i+1}+J_2\vec{S}_{i}\cdot\vec{S}_{i+2}\right),
\end{equation}
where $\vec{S}_i$ is the spin operator on the $i^{\text{th}}$-site. This model hosts a critical phase at $\frac{J_2}{J_1}=0.5$ in which it is known that the exact ground states are the two dimer states. In \cite{IMPS2010} the connection between fTNS an this model was established, and with our results we can now tell apart the two dimerized ground states based only on symmetry considerations, analogous to the treatment the AKLT model with MPS. Indeed, the first dimerized configuration, corresponds to the left of figure \ref{dimer} but for $N$ pairs of spins, in which we have seen that the dominant contribution carries the trivial representation on its virtual space. On the other hand, the opposite dimerized configuration, corresponding to the opposite pairing, will host a pair of spins on the edges that will carry a non-trivial representation which we identify with the topologically inequivalent groundstate.

Before concluding we should elaborate on what we mean by two fTNS corresponding to different critical SPT phases. While we can check which $SU(2)$-representation lies on the virtual space, a priori it could be possible that by redefining the parameters of the fTNS, such as the boundary functions $f_+(k),f_-(k)$ or its width $\Delta$, we could map to a different representation. A redefinition of the boundary functions alone is not enough to change the representation, since the new functions $\tilde{f}_+(k),\tilde{f}_-(k)$ must still be square-integrable. If we take a look at the last term of equation (\ref{amplitude}), we note that the spin representation is determined by the term $s_ie^{ikz_i}$. In order to change the representation, the new function would need to change this term, and since it is an exponential it would be impossible for it to remain in $\mathbb{L}^2(\mathbb{R})$. The only way in which this term could be absorbed would be if the width of the strip after the map $\tilde{\Delta}$ were to be such that the function remained integrable or if it simply changed the $s_i$ directly. However, there is no way of to redefine the functions in a way that simultaneously keeps the boundary functions in $\mathbb{L}^2(\mathbb{R})$ and keeps the sewing condition (\ref{sewing}) intact. Thus, the only way in which a fTNS can change its representation is with the value and position of the spin that it represents. Consequently, we call inequivalent two fTNS which describe different spin representations as they can not be mapped into one another by a redefinition of the virtual space.

\section{Conclusions}
In this paper we have derived the relation between the finite representation of SU(2) on the physical index of a fTNS and its corresponding representation as functional conformal charges on the virtual space. We have used this construction to identify the different topological properties of the two distinct ground states of the Majumdar-Ghosh point of the $J_1-J_2$ model. Our results are heavily inspired by the physics of the well-known MPS description of SPT phases which are analogous to our system. Several open directions can be taken now that this result is established. The construction of a conformal charge allows one to generate new models based on spin projectors, similar to \cite{WZW2011} for higher spins. It would also be possible to add new terms to the free boson action that would respect conformal symmetry in order to generate a new class of fTNS which would be the groundstate to a different critical Hamiltonian. Generalization of this result to the two-dimensional case presented in \cite{fTNS2021} would also be of high interest, given that the Kalmeyer-Laughlin is a prime example of chiral topological order, a state that has proven elusive for an exact description in terms of PEPS and whose symmetry study would be a great step towards having an exact tensor network-like description of gapped chiral 2D topological order. Another direction would be to study other simple CFT, such as the massless Majorana field of $c=\frac{1}{2}$ or the ghost systems, leading to yet another description of critical ground states and Hamiltonians. Understanding the exact connection between MERA and fTNS would also be of great interest. In general, deriving the exact connection between the CFT Hilbert space and fTNS would be an interesting task given that then one could study all the minimal models and other exactly solvable CFT using tensor network techniques. The connection between fTNS and cTNS would also be an interesting direction with the former being a potential boundary theory for the latter in the limit in which the spins become dense in space, thus providing a potential example beyond Gaussianity for cTNS. 

\section{Acknowledgements}
A. Gasull thanks David Stephen for countless discussions as well as acknowledges support from the International Max-Planck Research School for Quantum Science and Technology (IMPRS-QST). GS acknowledges financial support through the Spanish MINECO grant PID2021-127726NB-I00, the Comunidad de Madrid grant No. S2018/TCS-4342, the Centro de Excelencia Severo Ochoa Program SEV-2016-0597 and the CSIC Research Platform on Quantum Technologies PTI-001. JIC acknowledges funding from ERC Advanced Grant QUENOCOBA under the EU Horizon 2020 program (Grant Agreement No. 742102), and within the D-A-CH Lead-Agency Agreement through Project No. 414325145
(BEYOND C).

\bibliography{bibliography}
\clearpage
\onecolumngrid
\section*{Proof of the infinitesimal symmetry action} 
In this section we derive the rules of action for the conformal currents on a single strip of a fTNS, which would correspond to one of the two terms of \ref{algebrarelation}, where we are using that $q=\frac{s}{\sqrt{2}}$. We start by the action of the $H(z)$ current, whose charge is denoted by $Q^0$. First and foremost we need to choose a convention for the order in which we sew strips in situations in which we have more than two, which in this manuscript we choose to always sew from the lower boundary first, and then move upwards. In strip form, what we need to compute is
\begin{equation}
    \vcenter{\hbox{\includegraphics[width=0.25\linewidth]{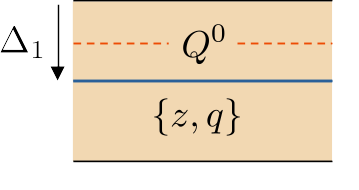}}} = (a).
    \label{appendix1}
\end{equation}

In equation form, equation (\ref{appendix1}) is written as
\begin{equation}
(a)=\int[\text{d}g]\frac{1}{2\pi i}\int_{\mathbb{R}}\text{d}z_1\mathcal{A}_{\Delta}[f_+,g,\{z,q\}]\lim_{q_1\rightarrow 0}\frac{\sqrt{2}}{q_1}\left(\partial_{z_1}\mathcal{A}_{\Delta_1}[g,f_-,\{z_1,q_1\}] - (z_1\Leftrightarrow\bar{z}_1)\right),
\end{equation}
where $\int_{\mathbb{R}}\text{d}z$ means the integration over $\mathbb{R}$ of the real part of $z_1$. Performing the sewing integral (\ref{ftnsewingbig}), one finds
\begin{equation}
    (a)=\frac{1}{2\pi i}\int_{\mathbb{R}}\text{d}z_1\lim_{q_1\rightarrow 0}\frac{\sqrt{2}}{q_1}\left[\partial_{z_1}\mathcal{A}_{\Delta_f}[f_+,f_-,\{z,q,z_1,q_1\}]- (z_1\Leftrightarrow\bar{z}_1)\right],
\end{equation}
where $\Delta_f=\Delta+\Delta_1$. Performing the derivative one obtains:
\begin{equation}
    \label{SM34}
    \begin{split}
         (a)=\frac{1}{2\pi i}\int_{\mathbb{R}}\text{d}z_1\lim_{q_1\rightarrow 0}\frac{\sqrt{2}}{q_1}&\left[\frac{i}{2}\int_{\mathbb{R}}\text{d}k\frac{ikq_1e^{ikz_1}}{\sinh{(\pi k \Delta_f)}}\left(e^{\pi k b_f}f_+(k)-e^{\pi k a_f}f_-(k)\right) - \right. \\ & \left.\frac{qq_1}{2\Delta_f}\coth{\left(\frac{z_1-z}{2\Delta_f}\right)}\left(\mu\sinh{\left(\frac{z_1-z}{2\Delta_f}\right)}\right)^{qq_1}\right]\mathcal{A}_{\Delta_f}[f_+,f_-,\{z,q,z_1,q_1\}]- (z_1\Leftrightarrow\bar{z}_1)
    \end{split}
\end{equation}
and we next treat both terms separately. Let us start with the first line, that is the integral
\begin{equation}
\frac{1}{2\pi i}\int_{\mathbb{R}}\text{d}z_1\frac{i}{\sqrt{2}}\int_{\mathbb{R}}\text{d}k\frac{ike^{ikz_1}}{\sinh{(\pi k \Delta_f)}}\left(e^{\pi k b_f}f_+(k)-e^{\pi k a_f}f_-(k)\right).
\end{equation}
We first swap the order of integration, that is we first perform the $z_1$-integral and then the $k$-integral. To be able to perform this change, it is enough to guarantee that the $k$-integral is convergent. We start by analyzing the behaviour of the integrand in the $k\rightarrow\pm\infty$ limits
\begin{equation}
    \begin{cases}
    & k\rightarrow+\infty \ \ \propto ke^{ikz_1-\pi k (\Delta_f-b_f)}f_+(k)-e^{ikz_1-\pi k (\Delta_f-a_f)}f_-(k) \rightarrow 0 \\ 
    &k\rightarrow-\infty \ \ \propto ke^{ikz_1+\pi k (\Delta_f+b_f)}f_+(k)-e^{ikz_1+\pi k (\Delta_f+a_f)}f_-(k)\rightarrow 0 ,
    \end{cases}
\end{equation}
where the decay to 0 in the limit is guaranteed because $\text{Im}(z_1)<\Delta_f=b_f-a_f$ and $f_{\pm}(k)$ are square integrable functions. The other potentially dangerous point is $k=0$, but the divergence is tamed by the power of $k$ in the numerator.  We can thus exchange the order of integrals and use the Dirac delta distribution to obtain : 
\begin{equation}
\frac{i}{\sqrt{2}}\int_{\mathbb{R}}\text{d}k\frac{k\delta{(k)}e^{-\pi k \text{Im}(z_1)}}{\sinh{(\pi k \Delta_f)}}\left(e^{\pi k b_f}f_+(k)-e^{\pi k a_f}f_-(k)\right).
\end{equation} 
All that is left is hence the evaluation of the $k$-integral by means of the Dirac distribution. In this case we must evaluate the limit $k\rightarrow0$
\begin{equation}
    \lim_{k\rightarrow0} \frac{ke^{-\pi k \text{Im}(z_1)}}{\sinh{(\pi k \Delta_f)}}\left(e^{\pi k b_f}f_+(k)-e^{\pi k a_f}f_-(k)\right) = \frac{1}{\pi\Delta_f}(f_+(0)-f_-(0)) = 0,
\end{equation}
where the last equality follows from the fact that the zero-mode is chosen to be the same amongst all the different sewing points on a state, as explained in (\cite{fTNS2021}). Thus, we have simplified (\ref{SM34}) down to
\begin{equation}
        \begin{split}
         (a)=\frac{1}{2\pi i}\int_{\mathbb{R}}\text{d}z_1\lim_{q_1\rightarrow 0}\frac{\sqrt{2}}{q_1}&\left[\frac{qq_1}{2\Delta_f}\coth{\left(\frac{z_1-z}{2\Delta_f}\right)}\left(\mu\sinh{\left(\frac{z_1-z}{2\Delta_f}\right)}\right)^{qq_1}\right]\mathcal{A}_{\Delta_f}[f_+,f_-,\{z,q,z_1,q_1\}]- (z_1\Leftrightarrow\bar{z}_1).
    \end{split}
\end{equation}
Now we can take the limit $q_1\rightarrow0$ obtaining
\begin{equation}
\label{SM40}
    (a)=\mathcal{A}_{\Delta_f}[f_+,f_-,\{z,q\}]\frac{q}{\sqrt{2}\Delta_f}\frac{1}{2\pi i}\int_{\mathbb{R}}\text{d}z_1\left[\coth{\left(\frac{z_1-z}{2\Delta_f}\right)} -\coth{\left(\frac{\bar{z}_1-z}{2\Delta_f}\right)}\right],
\end{equation}
where the limit removed the $z_1$ contribution of the functional $\mathcal{A}_{\Delta_f}$ and thus allows us to take it out of the integral. The remaining integral can be computed by residues calculus and it yields
\begin{equation}
   (a)=\frac{\sqrt{2}q\text{Im}(z_1)}{\pi\Delta_f} \mathcal{A}_{\Delta_f}[f_+,f_-,\{z,q\}].
\end{equation}
Finally we take the limit $\Delta_1\rightarrow 0$ to ensure the conformal dimension of the current, which in turn forces $\text{Im}(z_1)$ to be at the edge of the original strip, in this case the upper edge $\text{Im}(z_1)=\pi b$. This concludes this computation yielding
\begin{equation}
    (a)=\frac{\sqrt{2}qb}{\Delta} \mathcal{A}_{\Delta_f}[f_+,f_-,\{z,q\}].
    \label{axisofsigmaz}
\end{equation}
To conclude the computation of the commutator (\ref{sigmaz}), we now need to compute
\begin{equation}
        \vcenter{\hbox{\includegraphics[width=0.25\linewidth]{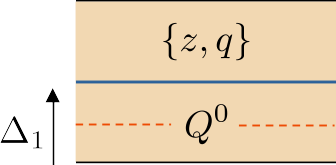}}} = (b),
    \label{appendix2}
\end{equation}

which is a computation that follows alongside the one we have just done. To see this, we can take a look at (\ref{SM34}) to see what the effect of sewing from the lower end would have. The main difference is that all the terms that depend on $z_1-z$ will now go as $z-z_1$ as well as the sign in front will change because of the derivative. Since the cotangent is an odd function, we recover (\ref{SM40}) at the end of the day. With this result, the commutator with $Q^0$ (\ref{sigmaz}) becomes
\begin{equation}
    (a)-(b)=\frac{\sqrt{2}qb}{\Delta} \mathcal{A}_{\Delta_f}[f_+,f_-,\{z,q\}] - \frac{\sqrt{2}qa}{\Delta} \mathcal{A}_{\Delta_f}[f_+,f_-,\{z,q\}] = \sqrt{2}q\mathcal{A}_{\Delta_f}[f_+,f_-,\{z,q\}],
\end{equation}
since $\Delta=b-a$. This is the expected action on single spin with the usual $\sigma_z$ operator, given that the charge is chosen to be $q=\frac{s}{\sqrt{2}}$. In strip form, this equality reads
\begin{equation}
    \vcenter{\hbox{\includegraphics[width=0.5\linewidth]{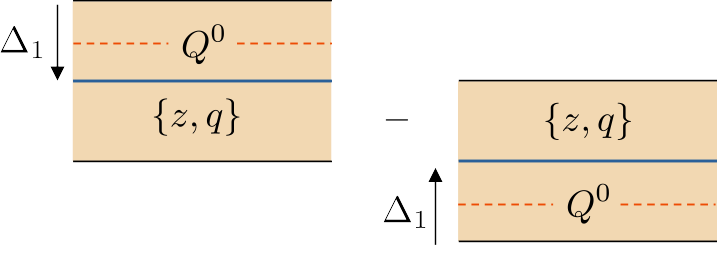}}} = s\mathcal{A}_{\Delta_f}[f_+,f_-,\{z,\frac{s}{\sqrt{2}}\}].
    \label{appendix3}
\end{equation}

We now turn our attention to the action of the lowering and raising currents $J^{\pm}(z)$, whose charges we will denote $Q^{\pm}$. As before, we start by the action on the upper edge of a strip which in strip form reads
\begin{equation}
    \vcenter{\hbox{\includegraphics[width=0.25\linewidth]{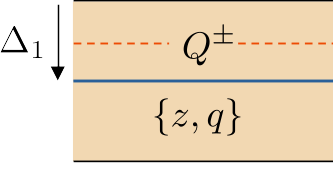}}} = (c),
    \label{appendix4}    
\end{equation}
or in equation form
\begin{equation}
    (c)=\int[\text{d}g]\frac{1}{2\pi i}\left(-\frac{\mu}{2}\right)\int_{\mathbb{R}}\text{d}z_1\mathcal{A}_{\Delta}[f_+,g,\{z,q\}]\mathcal{A}_{\Delta_1}[g,f_-,\{z_1,\pm\sqrt{2}\}] - (z_1\Leftrightarrow\bar{z}_1),
\end{equation}
where $-\frac{\mu}{2}$ ensures proper normalization at the end of the computation. We then perform the sewing and factorize what does not depend on $z_1$ outside of the integral to obtain
\begin{equation}
    (c)=\mathcal{A}_{\Delta}[f_+,f_-,\{z,q\}]\Delta^{-\frac{q_1^2}{2}}\frac{1}{2\pi i}\left(-\frac{\mu}{2}\right)\int_{\mathbb{R}}\text{d}z_1\left(\mu\sinh{\left(\frac{z_1-z}{2\Delta_f}\right)}\right)^{\pm\sqrt{2}q}\exp{\left(\frac{i}{2}\int_{\mathbb{R}}\text{d}k\frac{\pm\sqrt{2}e^{ikz_1}}{\sinh{(\pi k \Delta_f)}}\mathcal{C}(k)\right)} - (z_1\Leftrightarrow\bar{z}_1),
\end{equation}
where $\mathcal{C}(k) =\left(e^{\pi k b_f }\hat{f}_{+}(k)-e^{\pi k a_f}\hat{f}_{-}(k)\right) $ is a shorthand notation for the functional part of the boundary term. To tackle this integral we start by Taylor expanding the second exponential as
\begin{equation}
    \exp{\left(\frac{i}{2}\int_{\mathbb{R}}\text{d}k\frac{\pm\sqrt{2}e^{ikz_1}}{\sinh{(\pi k \Delta_f)}}\mathcal{C}(k)\right)} = \sum_{n=0}^{\infty}\int_{\mathbb{R}}\text{d}k_1...\text{d}k_n\left(\frac{i}{2}\right)^{n}\left(\pm\sqrt{2}\right)^n\prod^{n}_{m=1}\frac{\mathcal{C}(k_m)}{\sinh{(\pi k_m \Delta_f)}}e^{i\omega z_1},
\end{equation}
where $\omega = \sum_{l=1}^{m}k_l$. We can then again exchange the order of integration as both integrals are finite as we showed in the previous computation. Then,
\begin{equation}
\label{SM47}
    \begin{split}
    &(c)=\mathcal{A}_{\Delta}[f_+,f_-,\{z,q\}]\Delta^{-\frac{q_1^2}{2}}\frac{1}{2\pi i}\left(-\frac{\mu}{2}\right) \\ &\sum_{n=0}^{\infty}\int_{\mathbb{R}}\text{d}k_1...\text{d}k_n\left(\frac{i}{2}\right)^{n}\left(\pm\sqrt{2}\right)^n\prod^{n}_{m=1}\frac{\mathcal{C}(k_m)}{\sinh{(\pi k_m \Delta_f)}}\int_{\mathbb{R}}\text{d}z_1\left(\mu\sinh{\left(\frac{z_1-z}{2\Delta_f}\right)}\right)^{\pm\sqrt{2}q}e^{i\omega z_1} - (z_1\Leftrightarrow\bar{z}_1),
    \end{split}
\end{equation}
and thus we can apply residue calculus to the integral
\begin{equation}
\label{SM48}
    \int_{\mathbb{R}}\text{d}z_1\left(\mu\sinh{\left(\frac{z_1-z}{2\Delta_f}\right)}\right)^{\pm\sqrt{2}q}e^{i\omega z_1} - \left(\mu\sinh{\left(\frac{\bar{z}_1-z}{2\Delta_f}\right)}\right)^{\pm\sqrt{2}q}e^{i\omega \bar{z}_1}.
\end{equation}
In order to evaluate these integrals, we need to make a choice for both $q=\frac{s}{\sqrt{2}}$ as well as the sign of the current $J^{\pm}(z)$. We can start by first considering the case when we choose $J^{\pm}(z)$ and $q=\pm\frac{1}{\sqrt{2}}$, which corresponds to the case of annihilating the state by acting with the raising (lowering) operator on a state that is already the highest (lowest) element of the spin multiplet. In both of these cases the integral reads
\begin{equation}
        \mu\int_{\mathbb{R}}\text{d}z_1\sinh{\left(\frac{z_1-z}{2\Delta_f}\right)}e^{i\omega z_1} - \sinh{\left(\frac{\bar{z}_1-z}{2\Delta_f}\right)}e^{i\omega \bar{z}_1},
\end{equation}
and then we shall compute it by turning this integral into a contour integral. We start by more explicitly writing $z_1 = x+ iy$ and expanding the hyperbolic sines into exponentials as
\begin{equation}
    \frac{\mu}{2}\int_{\mathbb{R}}\text{d}xe^{\frac{x(i\omega+1)+y(i-\omega )-z}{2\Delta_f}}-e^{\frac{x(i\omega-1)-y(i+\omega)+z}{2\Delta_f}} - e^{\frac{x(i\omega+1)-y(i-\omega )-z}{2\Delta_f}}+e^{\frac{x(i\omega-1)+y(i+\omega)+z}{2\Delta_f}},
\end{equation}
that yields
\begin{equation}
    \mu\int_{\mathbb{R}}\text{d}x\sinh{\left(\frac{y(i-\omega )}{2\Delta_f}\right)}e^{\frac{x(i\omega+1)-z}{2\Delta_f}}+\sinh{\left(\frac{y(i+\omega )}{2\Delta_f}\right)}e^{\frac{x(i\omega-1)+z}{2\Delta_f}}.
\end{equation}
In order to ensure the convergence of these integrals, for $\omega>0$ we must extend the contour with a semicircle above the real axis, while for $\omega<0$ we must do so below the real axis. Special attention is required for the case $\omega=0$, where the integral reads
\begin{equation}
    \mu\sinh{\left(\frac{iy}{2\Delta_f}\right)}\int_{\mathbb{R}}\text{d}x\cosh{\left(\frac{x-z}{2\Delta_f}\right)},
\end{equation}
which is clearly divergent. However, this divergence gets exactly cancelled once the second term of the commutator is substracted. We can thus write
\begin{equation}
    \begin{split}
         & \Theta(\omega)\mu\oint_{\text{UHP}}\text{d}x\sinh{\left(\frac{y(i-\omega )}{2\Delta_f}\right)}e^{\frac{x(i\omega+1)-z}{2\Delta_f}}+\sinh{\left(\frac{y(i+\omega )}{2\Delta_f}\right)}e^{\frac{x(i\omega-1)+z}{2\Delta_f}}+ \\ 
         &\Theta(-\omega)\mu\oint_{\text{LHP}}\text{d}x\sinh{\left(\frac{y(i-\omega )}{2\Delta_f}\right)}e^{\frac{x(i\omega+1)-z}{2\Delta_f}}+\sinh{\left(\frac{y(i+\omega )}{2\Delta_f}\right)}e^{\frac{x(i\omega-1)+z}{2\Delta_f}} = 0,
    \end{split}
\end{equation}
where UHP/LHP stand for the sunrise contour going along the upper/lower half plane and $\Theta(\omega)$ is the step function. However, since this contours encircle no poles whatsoever, as the integrand has none, the result of this integral is simply zero. We thus find, that acting with the raising (lowering) current on the highest (lowest) states of a multiplet correctly sends them to zero. Of course, a sewing from below yields the same result.

We can now go back to (\ref{SM48}) and consider the case of $J^{\mp}(z)$ and $s=\pm$, which is the case in which we go from the higher to the lower state of the multiplet or viceversa. In this case (\ref{SM48}) reads
\begin{equation}
    \frac{1}{\mu}\int_{\mathbb{R}}\text{d}z_1\text{csch}{\left(\frac{z_1-z}{2\Delta_f}\right)}e^{i\omega z_1} - \text{csch}{\left(\frac{\bar{z}_1-z}{2\Delta_f}\right)}e^{i\omega \bar{z}_1},
\end{equation}
and performing a similar analysis as the previous one, we can write it as contours integrals as
\begin{equation}
\label{SM55}
        \begin{split}
         & \Theta(\omega)\frac{1}{\mu}\oint_{\text{UHP}}\text{d}z_1\text{csch}{\left(\frac{z_1-z}{2\Delta_f}\right)}e^{i\omega z_1} - \text{csch}{\left(\frac{\bar{z}_1-z}{2\Delta_f}\right)}e^{i\omega \bar{z}_1}+ \\ 
         &\Theta(-\omega)\frac{1}{\mu}\oint_{\text{LHP}}\text{d}z_1\text{csch}{\left(\frac{z_1-z}{2\Delta_f}\right)}e^{i\omega z_1} - \text{csch}{\left(\frac{\bar{z}_1-z}{2\Delta_f}\right)}e^{i\omega \bar{z}_1}.
    \end{split}
\end{equation}
Let us focus our attention first on the first line of (\ref{SM55}). If we write $z_1=x+iy$ as previously, then the poles of the first term are located at $x=z-iy+2\pi i n\Delta_f$ and the ones of the second term at $x=z+iy+2\pi i n\Delta_f$  for $n\in\mathbb{Z}$ . These are infinite towers of poles sitting in the imaginary axis, and the UHP contour encircles the poles corresponding to $n\in[1,\infty)$ for the first term and $n\in[0,\infty)$ for the second term since $z<iy$ as we are sewing from the upper edge. We thus evalute this integral using the residue theorem as
\begin{equation}
    \begin{split}
         & \Theta(\omega)\frac{1}{\mu}\oint_{\text{UHP}}\text{d}z_1\left[\text{csch}{\left(\frac{z_1-z}{2\Delta_f}\right)}e^{i\omega z_1} - \text{csch}{\left(\frac{\bar{z}_1-z}{2\Delta_f}\right)}e^{i\omega \bar{z}_1}\right]=\\ 
         &\Theta(\omega)\frac{1}{\mu}2\pi i\left[ \sum_{n=1}^{\infty}2\Delta_f(-1)^{n}e^{i\omega z +2\Delta_f\pi \omega n} -\sum_{n=0}^{\infty}2\Delta_f(-1)^{n}e^{i\omega z +2\Delta_f\pi \omega n} \right] = -\Theta(\omega)\frac{1}{\mu}2\pi i 2\Delta_fe^{i\omega z}.\\
    \end{split}
\end{equation}
Similarly for the second of line of (\ref{SM55}), the LHP contour encircles the poles corresponding to $n\in[0,-\infty)$ for the first term and to $n\in[-1,-\infty)$ for the second term for $n\in\mathbb{Z}$. Similarly, but with the contour now being counterclockwise the integral reads
\begin{equation}
        \begin{split}
         & \Theta(-\omega)\frac{1}{\mu}\oint_{\text{LHP}}\text{d}z_1\text{csch}{\left(\frac{z_1-z}{2\Delta_f}\right)}e^{i\omega z_1} - \text{csch}{\left(\frac{\bar{z}_1-z}{2\Delta_f}\right)}e^{i\omega \bar{z}_1}=\\ 
         &\Theta(-\omega)\frac{1}{\mu}(-2\pi i)\left[ \sum_{n=0}^{-\infty}2\Delta_f(-1)^{n}e^{i\omega z +2\Delta_f\pi \omega n} -\sum_{n=1}^{-\infty}2\Delta_f(-1)^{n}e^{i\omega z +2\Delta_f\pi \omega n} \right] = -\Theta(-\omega)\frac{1}{\mu}2\pi i 2\Delta_fe^{i\omega z}.\\
    \end{split}
\end{equation}
Collecting all the results we conclude
\begin{equation}
        \frac{1}{\mu}\int_{\mathbb{R}}\text{d}z_1\text{csch}{\left(\frac{z_1-z}{2\Delta_f}\right)}e^{i\omega z_1} - \text{csch}{\left(\frac{\bar{z}_1-z}{2\Delta_f}\right)}e^{i\omega \bar{z}_1} =-\frac{2\pi i}{\mu}2\Delta_fe^{i\omega z}.
\end{equation}
Inserting this result back into (\ref{SM47}) yields
\begin{equation}
    (c)=\mathcal{A}_{\Delta}[f_+,f_-,\{z,q\}]\frac{1}{\Delta_f}\frac{1}{2\pi i}\left(-\frac{\mu}{2}\right)\sum_{n=0}^{\infty}\int_{\mathbb{R}}\text{d}k_1...\text{d}k_n\left(\pm\frac{i}{2}\sqrt{2}\right)^{n}\prod^{n}_{m=1}\frac{\mathcal{C}(k_m)}{\sinh{(\pi k_m \Delta_f)}}(-\frac{2\pi i}{\mu})2\Delta_fe^{i\omega z},
\end{equation}
which allows us to collect the sum back into an exponential to finally write, after taking the $\Delta_1\rightarrow0$ limit
\begin{equation}
    (c)=\frac{1}{2}\mathcal{A}_{\Delta}[f_+,f_-,\{z,q\pm\sqrt{2}\}]\delta_{q,\mp\frac{1}{\sqrt{2}}}
\end{equation}
As before, sewing from the lower edge of the strip would again simply change the sign of the terms depending on $z_1-z$, and thus only change an overall minus sign. This allows us to finally reach the rules (\ref{sigmax}),(\ref{sigmay}),(\ref{sigmaz}) provided in the main text. Equation (\ref{sigmax}) reads in strip notation
\begin{equation}
\label{begin}
    \includegraphics[width=0.8\linewidth]{sigmax.png}
\end{equation}
or in equation form
\begin{align}
    &\lim_{\Delta_Q\rightarrow 0}\int[\mathcal{D}g]Q_{\Delta_Q}^x[f_+,g]\mathcal{A}_{\Delta}[g,f_-,\left\{z,\pm\frac{s}{\sqrt{2}}\right\}] = \frac{1}{2}\mathcal{A}_{\Delta}[f_+,f_-,\left\{z,\mp\frac{s}{\sqrt{2}}\right\}],\\
    &\lim_{\Delta_Q\rightarrow 0}\int[\mathcal{D}g]\mathcal{A}_{\Delta}[f_+,g,\left\{z,\pm\frac{s}{\sqrt{2}}\right\}]Q_{\Delta_Q}^x[g,f_-] = -\frac{1}{2}\mathcal{A}_{\Delta}[f_+,f_-,\left\{z,\mp\frac{s}{\sqrt{2}}\right\}].
\end{align}
Analogously, equation (\ref{sigmay}) reads in strip form
\begin{equation}
    \includegraphics[width=0.8\linewidth]{sigmay.png}
\end{equation}
or in equation form
\begin{align}
    &\lim_{\Delta_Q\rightarrow 0}\int[\mathcal{D}g]Q_{\Delta_Q}^y[f_+,g]\mathcal{A}_{\Delta}[g,f_-,\left\{z,\pm\frac{s}{\sqrt{2}}\right\}] = \pm\frac{i}{2}\mathcal{A}_{\Delta}[f_+,f_-,\left\{z,\mp\frac{s}{\sqrt{2}}\right\}],\\
    &\lim_{\Delta_Q\rightarrow 0}\int[\mathcal{D}g]\mathcal{A}_{\Delta}[f_+,g,\left\{z,\pm\frac{s}{\sqrt{2}}\right\}]Q_{\Delta_Q}^y[g,f_-] = \mp\frac{i}{2}\mathcal{A}_{\Delta}[f_+,f_-,\left\{z,\mp\frac{s}{\sqrt{2}}\right\}].
\end{align}
And finally, equation (\ref{sigmaz}) reads in strip form
\begin{equation}
    \includegraphics[width=0.8\linewidth]{sigmaz.png}
\end{equation}
or in equation form
\begin{align}
    &\lim_{\Delta_Q\rightarrow 0}\int[\mathcal{D}g]Q_{\Delta_Q}^z[f_+,g]\mathcal{A}_{\Delta}[g,f_-,\left\{z,\pm\frac{s}{\sqrt{2}}\right\}] = \pm\frac{s}{2}\mathcal{A}_{\Delta}[f_+,f_-,\left\{z,\pm\frac{s}{\sqrt{2}}\right\}],\\
    &\lim_{\Delta_Q\rightarrow 0}\int[\mathcal{D}g]\mathcal{A}_{\Delta}[f_+,g,\left\{z,\pm\frac{s}{\sqrt{2}}\right\}]Q_{\Delta_Q}^z[g,f_-] = \mp\frac{s}{2}\mathcal{A}_{\Delta}[f_+,f_-,\left\{z,\pm\frac{s}{\sqrt{2}}\right\}],
    \label{end}
\end{align}
where if we compare this last equation with equation (\ref{axisofsigmaz}) we see that we choose the limits of the strip to be $a=-\frac{\Delta}{2}$ and $b=\frac{\Delta}{2}$.
\section*{Proof of state invariance under the group action}
In this section we proof that a state that is described by a fTNS is invariant under the group action corresponding to the infinitesimal actions derived in the previous section. If we draw intuition from the usual SU(2) Lie group theory, the goal is to show that the state is invariant under a full rotation with angle $\theta$, and not only invariant with respect to the generators of said rotation. To proof this statement, the main formula of use is the following version of the renowned BCH formula :
\begin{equation}
e^{-i\theta X}Ye^{i\theta X}=Y+i\theta[Y,X]+\frac{(i\theta)^2}{2}[X,[X,Y]+...
\end{equation}
Since we have computed all the commutators in the previous section, it is easy to see the action of the whole exponential on a single strip, and they perfectly mirror the well known results for SU(2). If we write a state described by a FTN as
\begin{equation}
\label{SM63}
    |\psi\rangle = \sum_{s_1...s_N=\pm1}\int\mathcal{D}[f_1]...\mathcal{D}[f_N]\mathcal{A}_{f_1,f_2}^{s_1}...\mathcal{A}_{f_N,f_1}^{s_N}|s_1 ... s_N\rangle,
\end{equation}
and we act with the unitary matrix corresponding to a rotation around any of the axis $\alpha=x,y,z$, $U_i^{\alpha}(\theta)=\exp{(i\theta\sigma^{\alpha}_i)}$ on the $i^{\text{th}}$-spin, we can translate the action of this unitary onto the strip with the previously derived rules and the BCH formula. By moving the action to the virtual space we have
\begin{equation}
\begin{split}
    U_i^{\alpha}(\theta)|\psi\rangle &= \sum_{s_1...s_N=1}^d\int\mathcal{D}[f_1]...\mathcal{D}[f_N]\mathcal{A}_{f_1,f_2}^{s_1}...\mathcal{A}_{f_N,f_1}^{s_N}U_i^{\alpha}(\theta)|s_1 ... s_N\rangle \\&=  \sum_{s_1...s_N=1}^d\int\mathcal{D}[f_1]...\mathcal{D}[f_N]\mathcal{A}_{f_1,f_2}^{s_1}.\mathcal{A}_{f_i,f_{i+1}}^{U_i^{\alpha}(\theta)s_i}.\mathcal{A}_{f_N,f_1}^{s_N}|s_1 ... s_N\rangle,
    \end{split}
\end{equation}
where is what is meant by $\mathcal{A}_{f_i,f_{i+1}}^{U_i^{\alpha}(\theta)s_i}$ is that the unitary acts on the physical data of the $i^{\text{th}}$-strip, and thus it can be moved onto the virtual space by means of (\ref{MPSsymmetry}). Mathematically what we mean is
\begin{equation}
    \mathcal{A}_{f_i,f_{i+1}}^{U_i^{\alpha}(\theta)s_i} = \int\mathcal{D}[g]\mathcal{D}[f]\exp{(i\theta Q^{\alpha}[f_i,f])}\mathcal{A}_{f,g}^{s_i}\exp{(-i\theta Q^{\alpha}[g,f_{i+1}])},
\end{equation}
where now the BCH formula can be used to rewrite this in terms of commutators as
\begin{equation}
    \mathcal{A}_{f_i,f_{i+1}}^{U_i^{\alpha}(\theta)s_i} = \mathcal{A}_{f_i,f_{i+1}}^{s_i} + i\theta\left[Q^{\alpha},\mathcal{A}^{s_i}\right]_{f_i,f_{i+1}}+... \ .
\end{equation}
Before we proceed any further, it is important to recall that the common zero mode enforces the charge neutrality condition upon the closing of the fTNS, that means that any term of the superposition (\ref{SM63}) fulfills $\sum_{i=1}^{N}s_i=0$. We start by considering the charge $Q^{z}$ associated to the current $H(z)$, whose commutator acts as $\left[Q^{z},\mathcal{A}^{s_i}\right]= s_i \mathcal{A}_{f_i,f_{i+1}}^{s_i}$. Accordingly, the charges associated to the other generators act as $\left[Q^{x},\mathcal{A}^{\pm s_i}\right]=\mathcal{A}_{f_i,f_{i+1}}^{\mp s_i}$ and $\left[Q^{y},\mathcal{A}^{\pm s_i}\right]=\pm i \mathcal{A}_{f_i,f_{i+1}}^{\mp s_i}$, where all of these relations have been derived by repeated usage of the rules proofed in the last section. We can then re-sum the commutator expansion, similar to how one does it for Pauli matrices, and obtain
\begin{equation}
    \begin{split}
        & \mathcal{A}_{f_i,f_{i+1}}^{U_i^{z}(\theta)s_i} = e^{i\theta s_i}\mathcal{A}_{f_i,f_{i+1}}^{s_i}\\
        &\mathcal{A}_{f_i,f_{i+1}}^{U_i^{x}(\theta)(\pm s_i)} = \cos{\left(\frac{\theta}{2}\right)}\mathcal{A}_{f_i,f_{i+1}}^{\pm s_i}+i\sin{\left(\frac{\theta}{2}\right)}\mathcal{A}_{f_i,f_{i+1}}^{\mp s_i} \\
        &\mathcal{A}_{f_i,f_{i+1}}^{U_i^{y}(\theta)(\pm s_i)} = \cos{\left(\frac{\theta}{2}\right)}\mathcal{A}_{f_i,f_{i+1}}^{\pm s_i}\mp\sin{\left(\frac{\theta}{2}\right)}\mathcal{A}_{f_i,f_{i+1}}^{\mp s_i}.
    \end{split}
\end{equation}
Once we know the action of a full rotation on a strip, we can tackle the question of whether the full state is invariant under this operations. Clearly (\ref{SM63}) is not invariant under the action of a single unitary on a site, but only invariant under a unitary that acts on all spins simultaneously. We can easily see the invariance under rotations around the $z$-axis since
\begin{equation}
    \begin{split}
       &U_1^{z}(\theta)\otimes...U_N^{z}(\theta)|\psi\rangle = \sum_{s_1...s_N=\pm1}\int\mathcal{D}[f_1]...\mathcal{D}[f_N]\mathcal{A}_{f_1,f_2}^{U_1^{z}(\theta)s_1}...\mathcal{A}_{f_N,f_1}^{U_N^{z}(\theta)s_N}|s_1 ... s_N\rangle \\ &=\sum_{s_1...s_N=\pm1}e^{i\theta\sum_{i}s_i}\int\mathcal{D}[f_1]...\mathcal{D}[f_N]\mathcal{A}_{f_1,f_2}^{s_1}...\mathcal{A}_{f_N,f_1}^{s_N}|s_1 ... s_N\rangle=|\psi\rangle,
    \end{split}
\end{equation}
where the last equality follows from charge neutrality. For our concrete example of fTNs, the phase factor $\chi_{s_i}$ present in (\ref{HScoeffs}) are known collectively as the Marshall sign, which counts the number of "down"-spins on odd sites and gives a phase accordingly. This sign is the key to show invariance under rotations under any of the other two axis, and we show it for the $x$-axis by induction. Let us assume that $U_1^x(\theta)\otimes...\otimes U_{2n}^x(\theta)|\psi\rangle=|\psi\rangle$ for a state consisting of $n$ pairs of spins. Then, for a state consisting of $n+1$ pairs
\begin{equation}
    \begin{split}
  & U_1^x(\theta)\otimes...\otimes U_{2n}^x(\theta)\otimes U_{2n+1}^x(\theta)\otimes U_{2n+2}^x(\theta)|\psi\rangle= \\
  &U_1^x(\theta)\otimes...\otimes U_{2n}^x(\theta)\sum_{s_1...s_{2n}=\pm1}\sum_{s_{2n+1},s_{2n+2}=\pm1}\int\mathcal{D}[f_1]...\mathcal{D}[f_{2n+2}]\mathcal{A}_{f_1,f_2}^{s_1}...\mathcal{A}_{f_2n,f_1}^{s_{2n}}\mathcal{A}_{f_{2n+1},f_{2n+2}}^{U_{2n+1}^{x}(\theta)s_{2n+1}}\mathcal{A}_{f_{2n+2},f_{1}}^{U_{2n+2}^{x}(\theta)s_{2n+2}}|s_1 ... s_{2n+2}\rangle,
    \end{split}
\end{equation}
where we simply acted with the unitaries corresponding to the last pair. The action of these two unitaries yield
\begin{equation}
\label{SM70}
    \mathcal{A}_{f_{2n+1},f_{2n+2}}^{U_{2n+1}^{x}(\theta)\pm s_{2n+1}}\mathcal{A}_{f_{2n+2},f_{1}}^{U_{2n+2}^{x}(\theta)s_{2n+2}} = \left(\cos{\left(\frac{\theta}{2}\right)}\mathcal{A}_{f_{2n+1},f_{2n+2}}^{\pm s_{2n+1}}+i\sin{\left(\frac{\theta}{2}\right)}\mathcal{A}_{f_{2n+1},f_{2n+2}}^{\mp s_{2n+1}}\right)\left(\cos{\left(\frac{\theta}{2}\right)}\mathcal{A}_{f_{2n+2},f_{1}}^{\pm s_{2n+2}}+i\sin{\left(\frac{\theta}{2}\right)}\mathcal{A}_{f_{2n+2},f_{1}}^{\mp s_{2n+2}}\right),
\end{equation}
on the two functionals alone. Since the charge neutrality condition must be obeyed by all the terms of the superpositon of spins, only configurations that preserve it are able to contribute to the sum. That means, that for every term of the sum with fixed spin values $s_1...s_{2n}$, the two remaining spins can only be able to either flip its value or remain the same together. That means that we can then simplify (\ref{SM70}) to
\begin{equation}
    \cos{\left(\frac{\theta}{2}\right)}^2\mathcal{A}_{f_{2n+1},f_{2n+2}}^{\pm s_{2n+1}}\mathcal{A}_{f_{2n+2},f_{1}}^{\pm s_{2n+2}} -  \sin{\left(\frac{\theta}{2}\right)}^2\mathcal{A}_{f_{2n+1},f_{2n+2}}^{\mp s_{2n+1}}\mathcal{A}_{f_{2n+2},f_{1}}^{\mp s_{2n+2}}=\mathcal{A}_{f_{2n+1},f_{2n+2}}^{\pm s_{2n+1}}\mathcal{A}_{f_{2n+2},f_{1}}^{\pm s_{2n+2}},
\end{equation}
where the last equality follows from the fact that the first and second term are related by the Marshall sign. The Marshall sign always changes when two neighbouring spins swap values together, as that operation can only change the number of "down" spins that are siting at odd sites. This shows that we recover back the same state when we add an extra pair, as well as shows invariance of a single pair, thus concluding the proof.

\end{document}